\documentclass[12pt, onecolumn, draftcls]{IEEEtran}

\usepackage{epsf}
\usepackage{graphics}
\usepackage{times}
\usepackage{graphicx}
\usepackage{amsmath}
\usepackage{amssymb}
\usepackage{epsfig}
\usepackage{subfigure}

\usepackage{array}
\usepackage{float}

\newcommand{\comb}[2]{{}_{#1}\mathrm{C}_{#2}}

\begin{document}
\title{On the Phase Sequences and Permutation Functions in the SLM Scheme for OFDM-IM Systems}

\author{Kee-Hoon Kim
\thanks{The author is with the Department of Electronic Engineering, Soonchunhyang University, Asan 31538, Korea (e-mail: keehk85@gmail.com).}}
\maketitle

\begin{abstract}
In orthogonal frequency division
multiplexing with index modulation (OFDM-IM),
the active subcarriers can convey information bits by modulated
symbols as well as their indices. OFDM-IM has attracted a great deal of attention from researchers by virtue of high energy efficiency.
Unfortunately, OFDM-IM has inherited large peak-to-average power ratio (PAPR) of the classical OFDM signal, but there are few works dealing with it. Selected mapping (SLM) is a promising PAPR reduction technique that is distortionless, has good PAPR reducing capability, and can be easily adapted in OFDM-IM systems.
In SLM for OFDM-IM systems, the phase sequences rotate the OFDM-IM block in the frequency domain before the inverse fast Fourier transform (IFFT) is applied. Also, permutation in the frequency domain can be easily introduced before multiplication of phase sequences in SLM.
In this paper, we investigate the phase sequence set (PSS) and permutation functions in the SLM scheme for OFDM-IM systems; First, the efficiency of the permutation is analyzed as a function of the number of active subcarriers. Second, the optimal conditions for the PSS and permutation functions are derived, which is verified through simulations.
\end{abstract}


\section{Introduction}

Index modulation (IM) is an emerging technique that
exploits indices of active resources (such as spatial, time, and frequency resources) to convey information in
addition to conventional $M$-ary modulation symbols. The IM concept is applied to the orthogonal frequency division multiplexing (OFDM) \cite{frenger1999parallel,bacsar2013orthogonal}, which results
in a novel scheme termed as OFDM with index modulation
(OFDM-IM). Specifically, OFDM-IM activates only a subset of sub-carriers to carry data bits via both $M$-ary complex data symbols and active sub-carrier indices. Thus, OFDM-IM achieves higher energy efficiency and
reliability than the classical OFDM especially when it operates at a low bit rate. Therefore, OFDM-IM has attracted a great deal of attention from researchers as shown in recent surveys \cite{basar2017index,mao2018novel}.

As pointed out in \cite{ishikawa2016subcarrier}, OFDM-IM inherited the high peak-to-average power ratio (PAPR) problem from the classical OFDM. It is widely known that the high PAPR causes in-band distortion and out-of-band radiation when the OFDM-IM signal passes through high power amplifier (HPA). Lots of PAPR reduction methods have been researched in OFDM literatures \cite{han2005overview}. To solve the high PAPR problem in OFDM-IM, to borrow the methods for OFDM may work. However, those methods are not efficient because they do not consider the unique characteristic of OFDM-IM structure such as the presence of the inactive subcarriers \cite{zheng2017peak}.
At this point, there are only few works dealing with the large PAPR problem in OFDM-IM systems such as \cite{zheng2017peak,kim2019papr}, where the dither signals are used in the inactive subcarriers and thus induce in-band distortion.

Selected mapping (SLM) is a promising PAPR reduction technique that is distortionless, has good PAPR reducing capability, and can be easily adapted in OFDM-IM systems. The conventional SLM scheme is proposed in \cite{bauml1996reducing}. The SLM scheme has a probabilistic approach, where total $U$ alternative OFDM signals are generated and the best signal with the minimum PAPR is transmitted. To generate $U$ alternative OFDM signals, $U$ distinct phase sequences known to both transmitter and receiver are used. We denote the $U$ phase sequences as the phase sequence set (PSS) throughout this paper.
Since the SLM scheme has a disadvantage of high computational complexity, the many modified versions with low complexity have been proposed \cite{li2010novel,lim2005new,kim2016efficient}.

It is obvious that the PAPR reduction performance of the SLM scheme depends on the property of the used PSS. The authors in \cite{zhou2006optimality,lim2006phase,heo2009analysis,cheng2007improved} investigated the optimal condition for PSS in the SLM scheme for the classical OFDM system.
In summary, the element-wise multiplication of two phase sequences needs to be aperiodic. Therefore, the randomly generated PSS gives the sub-optimal PAPR reduction performance because the random sequence has aperiodicity with high probability. Also, the PSS using the rows of the cyclic Hadamard matrix constructed from an maximum length sequence (MLS) provides the near-optimal PAPR reduction performance. This is because the circular autocorrelation of an MLS is a Kronecker delta function.
Note that the PAPR reduction performance gap between the two solutions is small, but using the deterministic and systematic PSS has an advantage of performance guarantee and memory saving.

The conventional SLM scheme in \cite{bauml1996reducing} can be easily applied to the OFDM-IM systems, but the analytical results in \cite{zhou2006optimality,lim2006phase,heo2009analysis,cheng2007improved} cannot be naively applied to the OFDM-IM system because the OFDM-IM system has inactive subcarriers. Also, permutation procedure in the frequency domain can be adapted in the SLM scheme for OFDM-IM systems in order to make the subcarrier activation pattern (SAP) in the frequency domain diverse. This is because the envelope distribution of the OFDM-IM signal in the time domain depends on the SAP \cite{kim2016shift}. However, there is no work to analyze the permutation procedure in SLM.
To the authors' best knowledge, this is the first work to analyze the PSS and permutation functions in SLM for OFDM-IM systems.

In this paper, the SLM scheme with permutation is considered for OFDM-IM systems. Then, our contribution includes;
\begin{itemize}
  \item The efficiency of the permutation procedure is analyzed. Especially, the efficiency of permutation is derived as a function of the number of active subcarriers.
  \item The optimal conditions for PSS and the permutation functions are derived.
\end{itemize}

The rest of the paper can be summarized as follows. In
Section II, OFDM-IM, PAPR, and the conventional SLM scheme is reviewed. In
Section III, we introduce the conventional SLM with permutation procedure for OFDM-IM systems and investigate the efficiency of the permutation procedure. The optimal conditions for PSS and permutation functions are derived in Section IV. In Section V, we remark our contributions. Then, computer simulation results are presented to support the contributions in Section VI. Finally, Section VII concludes the paper.

\subsection{Notations}
Vectors are denoted by boldface letters $\mathbf{X}$ and its $i$-th element is denoted by $X(i)$. $\mathbf{X}_1\otimes \mathbf{X}_2$ denotes element-wise multiplication of two vectors $\mathbf{X}_1$ and $\mathbf{X}_2$. Also, $\comb{n}{k}$ means $n$ combination $k$, and $E[\cdot]$ denotes the statistical expectation. The complement of the set $\mathcal{I}$ is denoted by $\bar{\mathcal{I}}$, and the cardinality of the set $\mathcal{I}$ is denoted by $|\mathcal{I}|$.

\section{OFDM-IM, PAPR, and SLM}

\subsection{OFDM-IM and PAPR}
We consider an OFDM-IM system, where $N$ subcarriers are divided into
$G$ groups. Each group has $n$ subcarriers and $N=nG$. Unlike the conventional OFDM, not all the subcarriers are used for transmission in OFDM-IM. That is, only $k$ out of $n$ subcarriers are active in each group and clearly $k<n$. Then $p_1$ bits are transmitted by the SAP of each group. The number of possible SAPs in a group is $\comb{n}{k}$ and $p_1 = \lfloor \log_2 \comb{n}{k}\rfloor$.
Once the SAP is selected, $p_2 = k \log_2 M$ bits can be conveyed by $M$-ary modulation on the $k$ active subcarriers. Totally, $p = p_1 +p_2$ bits are transmitted per group.
Specifically, the $g$-th group is formed as
\begin{equation*}
  \mathbf{X}^{g} = [X^g(0)X^g(1)\cdots X^g(n-1)]
\end{equation*}
for $g=0,\cdots,G-1$, where only the $k$ elements in $\mathbf{X}^{g}$ are nonzero values from  $\mathcal{S}$, used signal constellation, and the others are set to zero.

After the formation of the $g$-th OFDM-IM group with
length $n$ for all $g$'s, they are concatenated \textit{in interleaving pattern} to
maximize the diversity gain \cite{xiao2014ofdm}. Throughout this paper, the interleaving pattern partitioning is assumed.
Then the OFDM-IM block in the frequency domain is
\begin{align*}
  \mathbf{X} &= [X(0) X(1)\cdots X(N-1)]\\
  &= [X^0(0)\cdots X^{G-1}(0)\cdots X^0(n-1)\cdots X^{G-1}(n-1)]
\end{align*}
or equivalently
\begin{equation*}
  X(G\cdot r + g) = X^g(r)
\end{equation*}
for $r=0,\cdots,n-1$ and $g=0,\cdots,G-1$.
We assume the average power of used constellation $\mathcal{S}$ is normalized and thus $E[|X(i)|^2]=k/n, i=0,\cdots,N-1$.

\begin{figure}
  \centering
  \includegraphics[width=.9\linewidth]{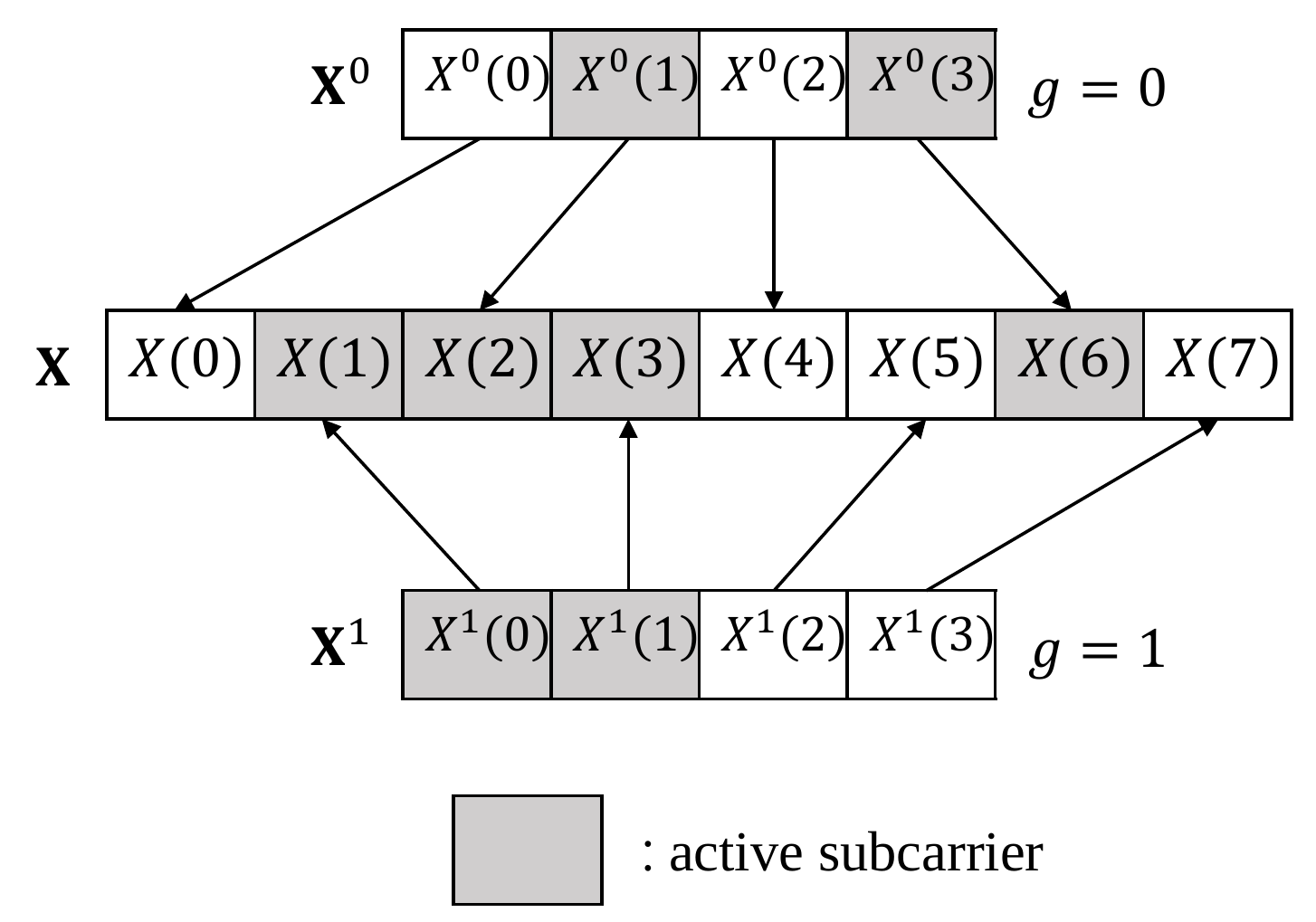}
  \caption{An example of an OFDM-IM block generation when $N=8$, $n=4$, $G=2$, and $k=2$ are used. Note that $\mathcal{I}=\{1,2,3,6\}$, $\mathcal{I}^0=\{2,6\}$, and $\mathcal{I}^1=\{1,3\}$.}\label{fig:X}
\end{figure}
Fig. \ref{fig:X} shows an example of an OFDM-IM block when $N=8$, $n=4$, and $k=2$ are used, where the OFDM-IM block $\mathbf{X}$ is constructed by concatenating $\mathbf{X}^0$ and $\mathbf{X}^1$ in interleaved pattern.
We define the SAP in the entire OFDM-IM block $\mathbf{X}$ as $\mathcal{I}$ and $|\mathcal{I}| = kG = K$. It is clear that $X(i) \in \mathcal{S}$ for $i \in \mathcal{I}$ and $X(i) =0$ for $i \in \bar{\mathcal{I}}$.
For example, in Fig. \ref{fig:X}, it is clear that $\mathcal{I}=\{1,2,3,6\}$ and $\bar{\mathcal{I}}=\{0,4,5,7\}$. Also, we define the $g$-th subset of $\mathcal{I}$ as $\mathcal{I}^g$, which is the set of indices related to the $g$-th group. In Fig. \ref{fig:X}, $\mathcal{I}^0=\{2,6\}$ and $\mathcal{I}^1=\{1,3\}$. Clearly, $\mathcal{I} = \bigcup_{g=0,\cdots,G-1}\mathcal{I}^g$.

After generation of the OFDM-IM block $\mathbf{X}$ in frequency domain, it is transformed into the OFDM-IM signal $\mathbf{x}$ in time domain by unitary inverse fast Fourier transform (IFFT) as
\begin{equation*}
  \mathbf{x} = \mbox{IFFT}\{\mathbf{X}\}
\end{equation*}
or
\begin{equation*}
  x(m) = \frac{1}{\sqrt{N}}\sum_{i=0}^{N-1}X(i)e^{\frac{j2\pi im}{N}}
\end{equation*}
for $m=0,\cdots,N-1$.

In the classical OFDM case, for a large $N$, elements of the OFDM signal are independent because of independent inputs to IFFT and approximately complex Gaussian distributed due to the central limit theorem. However, in OFDM-IM case, $x(0),\cdots,x(N-1)$ are approximately complex Gaussian distributed with zero mean, but they no longer have independency due to inactive subcarriers.
Also, the PAPR of $\mathbf{x}$ is defined as
\begin{equation*}
  \mbox{PAPR}(\mathbf{x}) = \frac{\max_{m=0,\cdots,N-1}|x(m)|^2}{E[|x(m)|^2]}
\end{equation*}
where $E[|x(m)|^2]=k/n$.

\subsection{The Conventional SLM Scheme}\label{sec:cSLM}
\begin{figure}
  \centering
  \includegraphics[width=.8\linewidth]{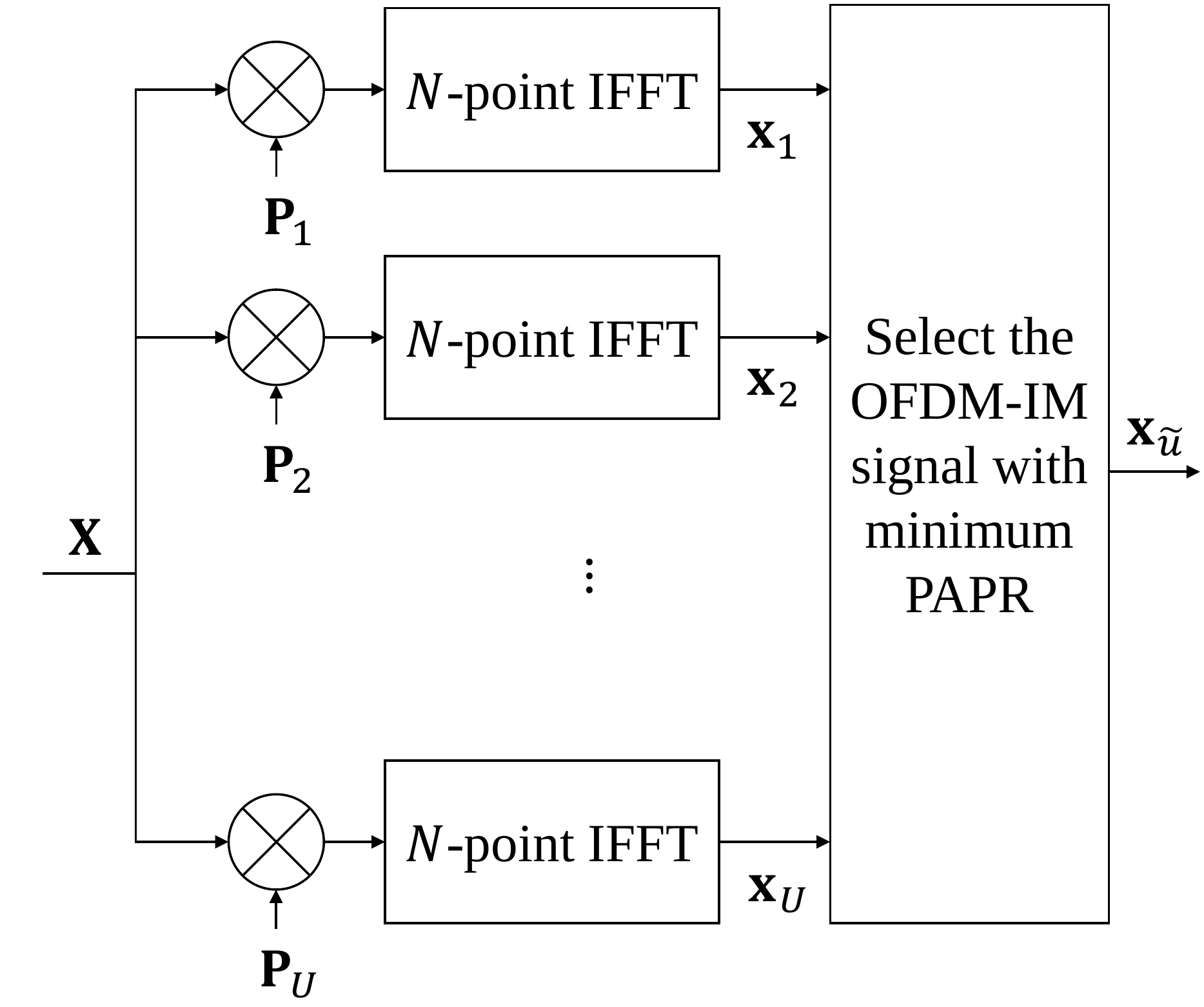}
  \caption{The conventional SLM scheme for OFDM-IM systems.}\label{fig:SLM}
\end{figure}

The conventional SLM scheme for OFDM systems is proposed in \cite{bauml1996reducing} and the conventional SLM for OFDM-IM systems is straightforward, which is depicted in Fig.\ref{fig:SLM}.
To generate $U$ alternative OFDM-IM signals, $U$ distinct phase sequences ${\mathbf P}_u,u=1,\cdots,U$ known to both transmitter and receiver are used, where ${\mathbf P}_u=[P_u(0)\cdots P_u(N-1)]$ with $P_u(i)=e^{j\phi_u(i)}$ and $\phi_u(i)\in [0,2\pi)$.
Specifically, the $u$-th alternative OFDM-IM signal is given by ${\mathbf x}_{u}= \mbox{IFFT}\{ \mathbf{P}_u \otimes\mathbf{X}\}$ for $u=1,\cdots,U$.
In this paper, the set $\{\mathbf{P}_1,\cdots,\mathbf{P}_U\}$ is referred to as PSS.
Then, the OFDM-IM signal with the minimum PAPR $\mathbf{x}_{\tilde{u}}$ is transmitted as
\begin{equation*}
\tilde{u}=\underset{u=1,\cdots,U}{\arg\min}\mbox{ PAPR}(\mathbf{x}_u).
\end{equation*}
Many modified versions of SLM have been proposed \cite{li2010novel,lim2005new,kim2016efficient}. They mainly focus on lowering the computational complexity because of high complexity of the SLM scheme, but the original methodology remains the same.

\section{The SLM scheme with permutation for OFDM-IM systems}
\begin{figure*}
  \centering
  \includegraphics[width=.7\linewidth]{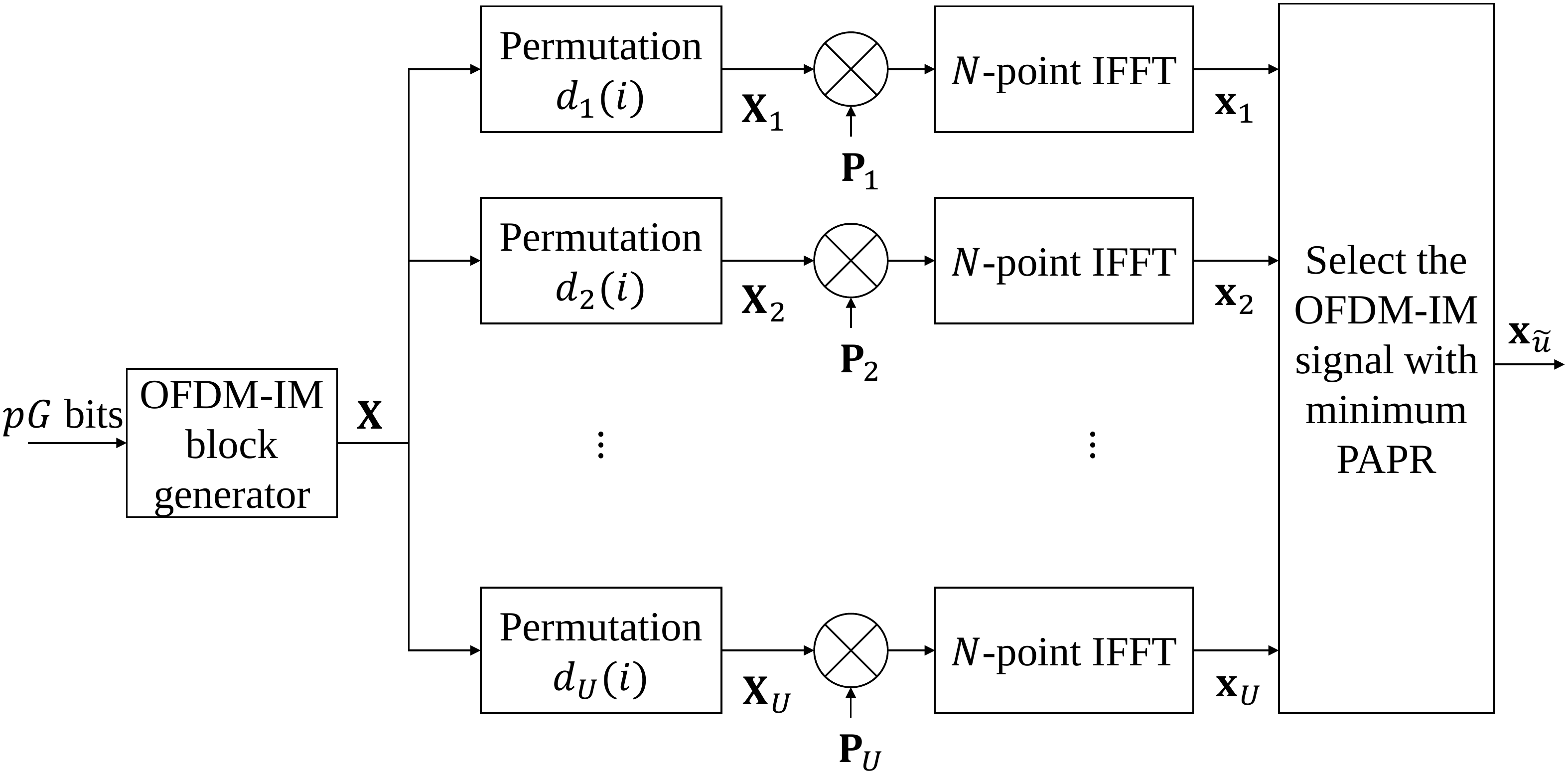}
  \caption{The SLM scheme with permutation for OFDM-IM systems.}\label{fig:blockdiagram}
\end{figure*}

Note that the SAP of an OFDM-IM block affects the envelope distribution of the corresponding OFDM-IM signal in the time domain and thus its PAPR \cite{kim2016shift}. Therefore, in OFDM-IM systems, the conventional SLM can be easily adapted with slight modification as in Fig. \ref{fig:blockdiagram}, where the permutation procedure followed by the multiplication of the phase sequence is newly introduced.

In particular, total $pG$-bits generate the OFDM-IM block ${\mathbf X}$, whose SAP is $\mathcal{I}$. Then, the OFDM-IM block ${\mathbf X}$ is permuted by the $u$-th permutation function $d_u(i),u=1,\cdots,U$, where the relation between ${\mathbf X}$ and its permuted version ${\mathbf X}_u=[X_u(0)\cdots X_u(N-1)]$ is
\begin{equation}\label{eq:perm}
  X_u(d_u(i)) = X(i)
\end{equation}
for $i=0,\cdots,N-1$. It is clear that the permutation should be done in each group \textit{individually}. This is because the elements in a group have to be distributed evenly to maximize the frequency diversity gain, even after the permutation.
It is obvious that the permutation procedure results in the change of the SAP of $\mathbf{X}$, and we denote the SAP of ${\mathbf X}_u$ as $\mathcal{I}_u,u=1,\cdots,U$.

From now on, the remaining procedures are the same as the conventional SLM in subSection \ref{sec:cSLM}. Specifically, the $u$-th permuted version $\mathbf{X}_u$ is multiplied by the $u$-th phase sequence ${\mathbf P}_u$. Then, these $U$ alternative OFDM-IM blocks $\mathbf{P}_u \otimes\mathbf{X}_u,u=1,\cdots,U$ in frequency domain are IFFTed to generate $U$ alternative OFDM-IM signals as
\begin{equation*}
  {\mathbf x}_{u}= \mbox{IFFT}\{ \mathbf{P}_u \otimes\mathbf{X}_u \}
\end{equation*}
for $u=1,\cdots,U$. Then, the PAPR values of them are calculated.
Finally, the alternative OFDM-IM signal ${\mathbf x}_{\tilde u}$ having the minimum PAPR is selected for transmission.

\subsection{Efficiency of Permutation in the SLM Scheme for OFDM-IM Systems}\label{sec:EP}
Now we investigate whether the permutation procedure in the SLM scheme in Fig. \ref{fig:blockdiagram} helps to improve the PAPR reduction performance or not.
If we use well-designed PSS, the $U$ alternative OFDM-IM signals in the SLM scheme become mutually independent. Then, complementary cumulative distribution function (CCDF) of PAPR of the transmitted OFDM-IM signal $\mathbf{x}_{\tilde{u}}$ is given by
\begin{align}\label{eq:CCDF}
  Pr\{\mbox{PAPR}(\mathbf{x}_{\tilde{u}}) >\gamma\}&=Pr\left\{\bigcap_{u=1}^{U}\mbox{PAPR}(\mathbf{x}_u)>\gamma\right\}\nonumber\\
  &=\prod_{u=1}^{U} Pr\{\mbox{PAPR}(\mathbf{x}_u)>\gamma\}.
\end{align}
Since $\mbox{PAPR}(\mathbf{x}_u)$ depends on the SAP $\mathcal{I}_u$, $Pr\{\mbox{PAPR}(\mathbf{x}_u)>\gamma\}$ is a function of $\gamma$ and  $\mathcal{I}_u$, and we rewrite it as
\begin{equation*}
  Pr\{\mbox{PAPR}(\mathbf{x}_u)>\gamma\}= F(\mathcal{I}_u,\gamma).
\end{equation*}

\subsubsection{Without Permutation}
Without permutation, the $U$ SAPs of $U$ alternative OFDM-IM signals are identical. That is, $\mathcal{I}_1 = \cdots = \mathcal{I}_U = \mathcal{I}$ and
we can consider them as random variables.
Then, the expectation of (\ref{eq:CCDF}) is
\begin{equation}\label{eq:woP}
  E[Pr\{\mbox{PAPR}(\mathbf{x}_{\tilde{u}})>\gamma\}]_{\mbox{woP}} = E[F(\mathcal{I},\gamma)^U].
\end{equation}

\subsubsection{With Permutation}
If we use random permutation, $\mathcal{I},\mathcal{I}_1,\cdots,\mathcal{I}_U$ becomes i.i.d. Then, the expectation of (\ref{eq:CCDF}) becomes
\begin{align}\label{eq:wP}
  E[Pr\{\mbox{PAPR}(\mathbf{x}_{\tilde{u}})>\gamma\}]_{\mbox{wP}}
  &=E[\prod_{u=1}^{U} F(\mathcal{I}_u,\gamma)]\nonumber\\
  &=\prod_{u=1}^{U} E[F(\mathcal{I}_u,\gamma)]\nonumber\\
  &=E[F(\mathcal{I},\gamma)]^U.
\end{align}
From (\ref{eq:woP}) and (\ref{eq:wP}), we have
\begin{equation}\label{eq:vs}
  E[Pr\{\mbox{PAPR}(\mathbf{x}_{\tilde{u}})>\gamma\}]_{\mbox{woP}}\geq E[Pr\{\mbox{PAPR}(\mathbf{x}_{\tilde{u}})>\gamma\}]_{\mbox{wP}}
\end{equation}
which means introducing the permutation procedure in SLM enhances the PAPR reduction performance.
Note that the difference between LHS and RHS in (\ref{eq:vs}) becomes large if $F(\mathcal{I},\gamma)$ has a large variance. In other words, as the random variable $\mathcal{I}$ varies, the diversity of the envelope distribution of OFDM-IM signals leads to the benefit of permutation in the SLM scheme.

\subsection{Quantifying the Efficiency of Permutation}\label{sec:Pandk}
In this subsection, we quantify the efficiency of permutation. First we investigate the envelope distribution of the OFDM-IM signal $\mathbf{x}$. The elements of $\mathbf{x}$ are correlated complex Gaussian with zero mean and their joint distribution are fully characterized by the correlation coefficient between the elements \cite{ochiai2001distribution}.

The correlation coefficient between $x(l)$ and $x(l+m)$ is
\begin{align}
  &\rho_{\mathcal{I}}(l,l+m)\nonumber\\
  &~= \frac{E[x(l) x^*(l+m)]-E[x(l)]E[x^*(l+m)]}{k/n}\nonumber \\
  &~= \frac{1}{N}\frac{E\left[\left(\sum_{i_1=0}^{N-1}X(i_1)e^{\frac{j2\pi i_1 l}{N}}\right)\left(\sum_{i_2=0}^{N-1}X(i_2)e^{\frac{j2\pi i_2 (l+m)}{N}}\right)^*\right]}{k/n} \nonumber\\
  &~=\frac{1}{Gk}\sum_{i\in \mathcal{I}}e^{-\frac{j2\pi i m}{N}}\label{eq:rho0}
\end{align}
where we use $E[X(i_1)X(i_2)^*]=1$ for $i_1= i_2\in \mathcal{I}$ and otherwise $E[X(i_1)X(i_2)^*]=0$.
It is noted that $\rho_{\mathcal{I}}(l,l+m)$ only depends on the index difference $m$ and can be rewritten as $\rho_{\mathcal{I}}(m)$. Also, from (\ref{eq:rho0}), we reconfirm that the envelope distribution of $\mathbf{x}$ is a function of $\mathcal{I}$.

Now we consider the SAP $\mathcal{I}$ as \textit{a random variable}. Note that, in a group, the number of possible SAPs is $\comb{n}{k}$. Since $\comb{n}{k} \geq 2^{p_1}$, redundancy $\comb{n}{k}-2^{p_1}$ SAPs are not used in a group. However, in this paper, we assume that all possible SAPs in a group are used with equal probability for simplicity. Clearly, the possible realization of $\mathcal{I}$ becomes $(\comb{n}{k})^G$.
Then, the correlation coefficient in (\ref{eq:rho0}) is
\begin{align*}
  &\rho_{\mathcal{I}}(m)\\
  &~=\frac{1}{Gk}\sum_{i\in \mathcal{I}}e^{-\frac{j2\pi i m}{N}} \nonumber\\
  &~=\frac{1}{Gk}\left(\sum_{i_0\in \mathcal{I}^0}e^{-\frac{j2\pi i_0 m}{N}}+\cdots+\sum_{i_{G-1}\in \mathcal{I}^{G-1}}e^{-\frac{j2\pi i_{G-1} m}{N}}\right) \nonumber\\
  &~=\frac{1}{Gk}\left(A_0 + e^{-\frac{j2\pi m}{N}}A_1 + \cdots+e^{-\frac{j2\pi(G-1)m}{N}}A_{G-1}\right)\nonumber
\end{align*}
where
\begin{equation}\label{eq:Ag}
  A_g = \alpha_{g}+\alpha_{g+G}e^{-\frac{j2\pi Gm}{N}}+\cdots+\alpha_{g+(n-1)G}e^{-\frac{j2\pi (n-1)Gm}{N}}
\end{equation}
and $\alpha_{i}$ is the random variable indicating the subcarrier activation as
\begin{equation}\label{eq:alpha}
  \alpha_{i} = \begin{cases}
                 1, & \mbox{if } X(i) \in \mathcal{S} \\
                 0, & \mbox{otherwise}
               \end{cases}
\end{equation}
for $i=0,\cdots,N-1$.
It is clear that $\alpha_{g}+\alpha_{g+G}+\cdots+\alpha_{g+(n-1)G} = k$ for $g=0,\cdots,G-1$ because there are $k$ active subcarriers per group.

As we mentioned, as the diversity of the envelope distribution of OFDM-IM signals increases, the efficiency of permutation increases.
Therefore, we can measure the efficiency of permutation by the variance of $\rho_{\mathcal{I}}(m)$.
Since $A_0,\cdots,A_{G-1}$ are i.i.d., the variance of $\rho_{\mathcal{I}}(m)$ is
\begin{align}
  \mbox{var}(\rho_{\mathcal{I}}(m))
  =& \frac{1}{G^2k^2}\left(\mbox{var}(A_0)+\cdots+\mbox{var}(A_{G-1})\right)\nonumber\\
  =& \frac{1}{Gk^2}\mbox{var}(A_0).\label{eq:varrho}
\end{align}
Now we investigate the variance of $A_0$.

\subsubsection{$m=0\mod n$}
From (\ref{eq:Ag}), we have
\begin{equation*}
  A_0 = \alpha_{0}+\alpha_{G}+\cdots+\alpha_{(n-1)G} =k
\end{equation*}
and
\begin{equation}\label{eq:var1}
  \mbox{var}(A_0)=0.
\end{equation}
\subsubsection{$m\neq0\mod n$}
Since $E[\alpha_i] = \frac{k}{n}$ for all $i$'s, from (\ref{eq:Ag}) we have
\begin{align}
  E[A_0] =& E[\alpha_{0}]+E[\alpha_{G}]e^{-\frac{j2\pi Gm}{N}}\nonumber\\
  &+\cdots+E[\alpha_{(n-1)G}]e^{-\frac{j2\pi (n-1)Gm}{N}}=0\label{eq:varA01}
\end{align}
and thus $|E[A_0]|^2=0$.
Also, from (\ref{eq:Ag}), we have
\begin{align}\label{eq:A02}
  &E[|A_0|^2]\nonumber\\
  &~= E\left[\left( \alpha_{0}+\alpha_{G}e^{\frac{j2\pi Gm}{N}}+\cdots+\alpha_{(n-1)G}e^{\frac{j2\pi (n-1)Gm}{N}}\right)\right.\nonumber\\
  &~~~\cdot\left.\left( \alpha_{0}+\alpha_{G}e^{-\frac{j2\pi Gm}{N}}+\cdots+\alpha_{(n-1)G}e^{-\frac{j2\pi (n-1)Gm}{N}}\right)\right]\nonumber\\
  &~=E\left[\sum_{i_0=0,G,\cdots,(n-1)G}\alpha_{i_0}^2e^{\frac{j2\pi i_0m}{N}}e^{-\frac{j2\pi i_0m}{N}}\right.\nonumber\\
  &~~~\left.+\sum_{i_1\neq i_2}\alpha_{i_1}\alpha_{i_2}e^{\frac{j2\pi i_1m}{N}}e^{-\frac{j2\pi i_2m}{N}}  \right]\nonumber\\
  &~=\frac{k}{n}\sum_{i_0=0,G,\cdots,(n-1)G}e^{\frac{j2\pi i_0m}{N}}e^{-\frac{j2\pi i_0m}{N}}\nonumber\\
  &~~~+\frac{k(k-1)}{n(n-1)}\sum_{i_1\neq i_2}e^{\frac{j2\pi i_1m}{N}}e^{-\frac{j2\pi i_2m}{N}}
\end{align}
where two following equations are used;
\begin{equation}\label{eq:alpha0}
  E[\alpha_{i_0}^2] = \frac{k}{n}
\end{equation}
for $i_0=0,\cdots,N-1$ and
\begin{equation}\label{eq:alpha1}
  E[\alpha_{i_1}\alpha_{i_2}] = \frac{\comb{n-2}{k-2}}{\comb{n}{k}} = \frac{k(k-1)}{n(n-1)}
\end{equation}
for $i_1 \neq i_2$ and $i_1,i_2\in\{0,G,\cdots,(n-1)G\}$. The equation (\ref{eq:alpha1}) can be derived by the fact $\alpha_{i_1}\alpha_{i_2}=1$ only if $\alpha_{i_1}=\alpha_{i_2}=1$.

Then, we can rewrite (\ref{eq:A02}) as
\begin{align}
  &E[|A_0|^2]\nonumber\\
  &~=\left(\frac{k}{n}-\frac{k(k-1)}{n(n-1)}\right)\sum_{i_0=0,G,\cdots,(n-1)G}e^{\frac{j2\pi i_0m}{N}}e^{-\frac{j2\pi i_0m}{N}}\nonumber\\
  &~~~+\frac{k(k-1)}{n(n-1)}\left(\sum_{i_0=0,G,\cdots,(n-1)G}e^{\frac{j2\pi i_0m}{N}}e^{-\frac{j2\pi i_0m}{N}}\right.\nonumber\\
  &~~~\left.+\sum_{i_1\neq i_2}e^{\frac{j2\pi i_1m}{N}}e^{-\frac{j2\pi i_2m}{N}}\right)\nonumber\\
  &~=\frac{k(n-k)}{n-1}\nonumber\\
  &~~~+\frac{k(k-1)}{n(n-1)}\left(e^{\frac{j2\pi0m}{N}}+e^{\frac{j2\pi Gm}{N}}+\cdots+e^{\frac{j2\pi(n-1)Gm}{N}}\right)\nonumber\\
  &~~~\cdot\left(e^{-\frac{j2\pi0m}{N}}+e^{-\frac{j2\pi Gm}{N}}+\cdots+e^{-\frac{j2\pi(n-1)Gm}{N}}\right)\label{eq:A02'}\\
  &~=\frac{k(n-k)}{n-1}\label{eq:varA02}
\end{align}
where the second term in (\ref{eq:A02'}) becomes zero because $m\neq0\mod n$.
Then, from (\ref{eq:varA01}) and (\ref{eq:varA02}), we have
\begin{equation}\label{eq:var2}
  \mbox{var}(A_0)=\frac{k(n-k)}{n-1}.
\end{equation}

In conclusion, plugging (\ref{eq:var1}) and (\ref{eq:var2}) into (\ref{eq:varrho}) gives
\begin{equation}\label{eq:rhocase}
  \mbox{var}(\rho_{\mathcal{I}}(m))=\begin{cases}
                                      \frac{1}{N}\frac{n}{n-1}\left(\frac{n}{k}-1\right), & m\neq0\mod n \\
                                      0, & m = 0\mod n
                                    \end{cases}
\end{equation}
where the zero variance means that the value of $\rho_{\mathcal{I}}(m)$ is a constant value when $m = 0\mod n$.
Note that as $k$ decreases, the variance of $\rho_{\mathcal{I}}(m)$ increases, which means the gap between LHS and RHS terms in (\ref{eq:vs}) becomes larger as $k$ decreases. In conclusion, the efficiency of the permutation increases as $k$ decreases.

\section{The Optimal Condition for PSS and Permutation Functions}
In this section, we will derive the optimal condition for phase sequences and permutation functions in the SLM scheme for OFDM-IM systems.
For simplicity, we will investigate the conditions for two alternative OFDM-IM signals $\mathbf{x}_1$ and $\mathbf{x}_2$. First, we consider the optimal condition for PSS. As seen in (\ref{eq:CCDF}), the phase sequences should make these two alternative OFDM-IM signals independent. That is, the set $\{x_1(l)\}_{l=0}^{N-1}$ and $\{x_2(m)\}_{m=0}^{N-1}$ should be mutually independent.
Since elements of an OFDM-IM signal are correlated complex Gaussian with zero mean due to the central limit theorem \cite{ochiai2001distribution}, zero covariance between them induces the independency. The covariance between $x_1 (l)$ and $x_2(m)$ is given by
\begin{align*}
  &\mbox{cov}(x_1 (l) x_2 (m) )\nonumber\\
  &~= E[x_1 (l) x_2^* (m) ]-E[x_1 (l)]E[x_2^* (m)] \nonumber\\
  &~= \frac{1}{N}E\left[\left(\sum_{i_1=0}^{N-1}P_1(i_1)X_1(i_1)e^{\frac{j2\pi i_1 l}{N}}\right)\right.\nonumber\\
  &~~~\cdot\left.\left(\sum_{i_2=0}^{N-1}P_2(i_2)X_2(i_2)e^{\frac{j2\pi i_2 m}{N}}\right)^*\right] \nonumber\\
  &~= \frac{1}{N}E\left[\sum_{i_1=0}^{N-1}\sum_{i_2=0}^{N-1}P_1(i_1)P_2^*(i_2)X_1(i_1)X_2^*(i_2)\right.\nonumber\\
  &~~~\left.e^{\frac{j2\pi i_1 l}{N}}e^{-\frac{j2\pi i_2 m}{N}}\right].
\end{align*}
From (\ref{eq:perm}), we have
\begin{align*}
  &E[X_1(i_1)X_2^*(i_2)]\\
  &~=
  \begin{cases}
  1, & \mbox{if $i_1 = d_1(i)$ and $i_2 = d_2(i)$ for all $i\in \mathcal{I}$}\\
  0, & \mbox{otherwise}.
  \end{cases}
\end{align*}
Then, $|\mbox{cov}(x_1 (l) x_2 (m) )|$ becomes
\begin{align}\label{eq:abscov}
  &|\mbox{cov}(x_1 (l) x_2 (m) )|\nonumber\\
  &~= \frac{1}{N}\left|\sum_{i\in \mathcal{I}}P_1(d_1(i))P_2^*(d_2(i))e^{\frac{j2\pi d_1(i) l}{N}}e^{-\frac{j2\pi d_2(i)m}{N}}\right|.
\end{align}

\subsection{Optimal Condition for PSS without Permutation}\label{sec:psswoP}
As we analyzed, the efficiency of permutation is small if $k$ is large, which means the permutation may not used in the SLM scheme for large $k$ compared to $n$. Therefore, we first consider the case of no permutation is used in the SLM scheme. In this case, clearly we have $d_1(i)=d_2(i)=i$ and thus (\ref{eq:abscov}) becomes
\begin{align}
  |\mbox{cov}(x_1 (l) x_2 (m) )|
  &= \frac{1}{N}\left|\sum_{i\in \mathcal{I}}P_1(i)P_2^*(i)e^{\frac{j2\pi i (l-m)}{N}}\right|\nonumber\\
  &= \frac{1}{\sqrt{N}}\left|\underset{i\in \mathcal{I}}{\mbox{IDFT}}\{P_1(i)P_2^*(i)\}\right|\label{eq:confornoP}
\end{align}
where $\underset{i\in \mathcal{I}}{\mbox{IDFT}}\{\cdot\}$ denotes the unitary inverse discrete Fourier transform (IDFT) whose participating indices of input are the elements in $\mathcal{I}$ only. To boost the PAPR reduction performance of the SLM scheme, the sequence (\ref{eq:confornoP}) should be as low as possible. In other words, the ``punctured'' version of the sequence $\mathbf{P}_1\otimes \mathbf{P}_2^*$ has to be aperiodic.

To satisfy the condition, one of the sub-optimal solutions is to use the randomly generated PSS because random sequence has aperiodicity with high probability. Also, as a near-optimal and deterministic solution is to use the rows of the cyclic Hadamard matrix constructed from an MLS for $\mathbf{P}_u$'s for $u=1,\cdots,U$, which is the same solution in \cite{lim2006phase} for the classical OFDM. The reason is as follows;

Suppose that we use the PSS satisfying
\begin{equation*}
  \frac{1}{N}\left|\sum_{i=0}^{N-1}P_1(i)P_2^*(i)e^{\frac{j2\pi i m}{N}}\right|\leq c
\end{equation*}
where $c$ is a constant.
Since $|\bar{\mathcal{I}}|=N-K$, we also have
\begin{equation*}
  \frac{1}{N}\left|\sum_{i\in \bar{\mathcal{I}}}P_1(i)P_2^*(i)e^{\frac{j2\pi i m}{N}}\right|\leq \frac{N-K}{N}.
\end{equation*}
Using the triangular inequality for (\ref{eq:confornoP}), we have
\begin{align}
  |\mbox{cov}(x_1 (l) x_2 (m) )| \leq& \frac{1}{N}\left|\sum_{i=0}^{N-1}P_1(i)P_2^*(i)e^{\frac{j2\pi i m}{N}}\right|\nonumber\\
  &+ \frac{1}{N}\left|\sum_{i\in \bar{\mathcal{I}}}P_1(i)P_2^*(i)e^{\frac{j2\pi i m}{N}}\right|\nonumber\\
  \leq& c+\frac{N-K}{N} = c + 1-\frac{k}{n}.\label{eq:bound}
\end{align}
Therefore, without permutation, using the rows of the cyclic Hadamard matrix for $\mathbf{P}_u$ provides the small $c$ and thus small upper bound in (\ref{eq:bound}).

\subsection{Optimal Condition for PSS with Permutation}\label{sec:PSSwithP}
Now we consider the SLM scheme with permutation procedure as in Fig. \ref{fig:blockdiagram}. Using the substitution $i'=d_2(i)$ and $i = d_2^{-1}(i')$,
(\ref{eq:abscov}) is rewritten as
\begin{align}\label{eq:opconwP}
  &|\mbox{cov}(x_1 (l) x_2 (m) )|\nonumber\\
  &~= \frac{1}{N}\left|\sum_{i'\in \mathcal{I}'}P_1(d_1(d_2^{-1}(i')))P_2^*(i')e^{\frac{j2\pi d_1(d_2^{-1}(i')) l}{N}}e^{-\frac{j2\pi i'm}{N}}\right|\nonumber\\
  &~= \frac{1}{\sqrt{N}}\left|\underset{i'\in \mathcal{I}'}{\mbox{DFT}}\left\{P_1(d_1(d_2^{-1}(i')))P_2^*(i')e^{\frac{j2\pi d_1(d_2^{-1}(i')) l}{N}}\right\}\right|
\end{align}
where $\mathcal{I}'=\{d_2(i)|i\in\mathcal{I} \}$ and $\underset{i'\in \mathcal{I}'}{\mbox{DFT}}\{\cdot\}$ denotes the unitary discrete Fourier transform (DFT) whose participating indices of the input are the elements in $\mathcal{I}'$ only.
Therefore, the optimal condition for the PSS is as follows;
For all $l$'s, the input sequence of DFT in (\ref{eq:opconwP}) should be aperiodic.
Since the permutation functions affect that sequence and there are totally $N$ sequences to be considered because of $l=0,\cdots,N-1$, it is hard to find the systematic PSS satisfying this property. Therefore, we defer it as a future work.
Instead, we may use randomly generated PSS as a sub-optimal solution.
Then, the input of DFT in (\ref{eq:opconwP}) for all $l$'s become random sequences regardless of the permutation functions, and the random sequence is aperiodic with high probability.

\subsection{Optimal Condition for Permutation Functions $d_u(i)$}\label{sec:conforP}
In this subsection, we will investigate the optimal condition for permutation functions in the SLM scheme. As we mentioned, permutation has an effect on changing the SAP and thus $\rho_{\mathcal{I}}(m)$. Therefore, we investigate the covariance between $\rho_{\mathcal{I}_1}(l)$ and $\rho_{\mathcal{I}_2}(m)$, where $\mathcal{I}_1$ and $\mathcal{I}_2$ are the SAPs of the first and second alternative OFDM-IM signals, respectively. From (\ref{eq:rho0}) we have
\begin{align}
  &G^2k^2\mbox{cov}(\rho_{\mathcal{I}_1}(l)\rho_{\mathcal{I}_2}(m))\nonumber\\
  &~= E\left[\left(\sum_{i_1\in \mathcal{I}_1}e^{-\frac{j2\pi i_1 l}{N}}\right)\left(\sum_{i_2\in \mathcal{I}_2}e^{-\frac{j2\pi i_2 m}{N}}\right)^*\right]\nonumber\\
  &~~~-E\left[\left(\sum_{i_1\in \mathcal{I}_1}e^{-\frac{j2\pi i_1 l}{N}}\right)\right]E\left[\left(\sum_{i_2\in \mathcal{I}_2}e^{-\frac{j2\pi i_2 m}{N}}\right)^*\right]\label{eq:exey}
\end{align}
where the second term in (\ref{eq:exey}) is zero except when $l=m=0$ and can be negligible because of (\ref{eq:rhocase}). Then, the equation in (\ref{eq:exey}) is rewritten as
\begin{align}
  &G^2k^2\mbox{cov}(\rho_{\mathcal{I}_1}(l)\rho_{\mathcal{I}_2}(m))\nonumber\\
  &~=E\left[\left(\sum_{i_{1,0}\in \mathcal{I}_1^0}e^{-\frac{j2\pi i_{1,0} l}{N}}+\cdots+\sum_{i_{1,G-1}\in \mathcal{I}_1^{G-1}}e^{-\frac{j2\pi i_{1,G-1} l}{N}}\right)\right.\nonumber\\
  &~~~\cdot\left.\left(\sum_{i_{2,0}\in \mathcal{I}_2^0}e^{\frac{j2\pi i_{2,0} m}{N}}+\cdots+\sum_{i_{2,G-1}\in \mathcal{I}_2^{G-1}}e^{\frac{j2\pi i_{2,G-1} m}{N}}\right)\right]\label{eq:exy}
\end{align}
where $\mathcal{I}_u^g,u=1,2,g=0,\cdots,G-1$ is the subset of $\mathcal{I}_u$, consists of the indices related to the $g$-th group only.

Since two distinct groups are independent, the cross terms in (\ref{eq:exy}) is
\begin{align*}
  &E\left[\sum_{i_{1,g_1}\in \mathcal{I}_1^{g_1}}e^{-\frac{j2\pi i_{1,g_1}l}{N}}\right]E\left[\sum_{i_{2,g_2}\in \mathcal{I}_2^{g_2}}e^{\frac{j2\pi i_{2,g_2}m}{N}}\right]\\
  &~=\frac{k^2}{n^2}\sum_{i_{1,g_1}=g_1,G+g_1,\cdots,(n-1)G+g_1}e^{-\frac{j2\pi i_{1,g_1}l}{N}}\\
  &~~~\cdot\sum_{i_{2,g_2}=g_2,G+g_2,\cdots,(n-1)G+g_2}e^{\frac{j2\pi i_{2,g_2}m}{N}}
\end{align*}
for $0\leq g_1\neq g_2 \leq G-1$, where this term is zero except when $l=0\mod n$ and $m=0 \mod n$. Then the cross terms in (\ref{eq:exy}) is negligible because of (\ref{eq:rhocase}).
Then, we can rewrite (\ref{eq:exy}) as
\begin{align}
  &G^2k^2\mbox{cov}(\rho_{\mathcal{I}_1}(l)\rho_{\mathcal{I}_2}(m))\nonumber\\
  &~= \sum_{g=0}^{G-1}\underbrace{E\left[\sum_{i_{1,g}\in \mathcal{I}_1^g}e^{-\frac{j2\pi i_{1,g} l}{N}}\sum_{i_{2,g}\in \mathcal{I}_2^g}e^{\frac{j2\pi i_{2,g} m}{N}}\right]}_{B_g}.\label{eq:sigmaB}
\end{align}
Using (\ref{eq:perm}) and (\ref{eq:alpha}), $B_g$ can be rewritten as
\begin{align*}
  &B_g\nonumber\\
  &~=E\left[\left(\alpha_g e^{-\frac{j2\pi d_1(g) l}{N}}+\alpha_{G+g} e^{-\frac{j2\pi d_1(G+g) l}{N}}\right.\right.\\
  &~~~+\cdots+\left.\alpha_{(n-1)G+g} e^{-\frac{j2\pi d_1((n-1)G+g) l}{N}}\right)\nonumber\\
  &~~~\cdot\left(\alpha_g e^{\frac{j2\pi d_2(g) m}{N}}+\alpha_{G+g} e^{\frac{j2\pi d_2(G+g) m}{N}}\right.\\
  &~~~+\cdots+\left.\left.\alpha_{(n-1)G+g} e^{\frac{j2\pi d_2((n-1)G+g) m}{N}}\right)\right].\nonumber
\end{align*}

Since the permutation is done in each group individually, we have
\begin{align}\label{eq:set2}
  &\{d_u(g),d_u(G+g),\cdots,d_u((n-1)G+g)\}\nonumber\\
  &~=\{g,G+g,\cdots,(n-1)G+g\}.
\end{align}
Using (\ref{eq:alpha0}), (\ref{eq:alpha1}), and (\ref{eq:set2}), $B_g$ becomes
\begin{align}
  &B_g\nonumber\\
  &~=\frac{k(k-1)}{n(n-1)}\left(e^{-\frac{j2\pi gl}{N}}+e^{-\frac{j2\pi (G+g)l}{N}}
  +\cdots+e^{-\frac{j2\pi((n-1)G +g)l}{N}}\right)\nonumber\\
  &~~~\cdot\left(e^{\frac{j2\pi gm}{N}}+e^{\frac{j2\pi(G+g) m}{N}}+\cdots+e^{\frac{j2\pi((n-1)G+g)m}{N}}\right)\nonumber\\
  &~~~+\underbrace{\left(\frac{k}{n}-\frac{k(k-1)}{n(n-1)}\right)\sum_{i=0,G,\cdots,(n-1)G}e^{\frac{j2\pi(d_2(i+g)m-d_1(i+g)l)}{N}}}_{D_g}\nonumber\\
  &~=C_g + D_g.\label{eq:second}
\end{align}
The term $C_g$ in (\ref{eq:second}) becomes
\begin{equation}\label{eq:Cg}
  C_g = \begin{cases}
          \frac{k(k-1)}{n(n-1)}n^2, & \mbox{if } l = 0 \mod n \mbox{ and } m = 0 \mod n\\
          0, & \mbox{otherwise}.
        \end{cases}
\end{equation}
From (\ref{eq:Cg}) and (\ref{eq:rhocase}), $C_g$ can be negligible. Therefore, we only investigate $D_g$ in (\ref{eq:second}). Then, (\ref{eq:sigmaB}) becomes
\begin{align}
  &G^2k^2\mbox{cov}(\rho_{\mathcal{I}_1}(l)\rho_{\mathcal{I}_2}(m))\nonumber\\
  &~=\sum_{g=0}^{G-1}D_g\nonumber\\
  &~=\frac{k(n-k)}{n(n-1)}\sum_{g=0}^{G-1}\sum_{i=0,G,\cdots,(n-1)G}e^{\frac{j2\pi(d_2(i+g)m-d_1(i+g)l)}{N}}\nonumber\\
  &~=\frac{k(n-k)}{n(n-1)}\sum_{i=0}^{N-1}e^{\frac{j2\pi(d_2(i)m-d_1(i)l)}{N}}\nonumber\\
  &~=\frac{k(n-k)}{n(n-1)}\sum_{i'=0}^{N-1}e^{\frac{j2\pi(i'm-d_1(d_2^{-1}(i'))l)}{N}}\nonumber\\
  &~=\frac{k(n-k)}{n(n-1)}\sqrt{N}\underset{i'=0,\cdots,N-1}{\mbox{IDFT}}\left\{e^{-\frac{j2\pi d_1(d_2^{-1}(i'))l}{N}}  \right\}\label{eq:permdesign}
\end{align}
where the substitution $i'=d_2(i)$ and $d_2^{-1}(i')=i$ are used.

In conclusion, the optimal condition for the permutation functions is that (\ref{eq:permdesign}) should have low envelope fluctuation or variance for all $l$'s. In other words, the input sequence of IDFT in (\ref{eq:permdesign}) should be aperiodic for all $l$'s. Clearly, random permutation could be the sub-optimal solution.

\section{Remarks}
We remark the contributions of our work before simulation results are presented. For the conventional SLM scheme with permutation in OFDM-IM systems;
\begin{itemize}
  \item We derived the efficiency of permutation and found that the efficiency of the permutation increases as $k$ decreases. (subSection \ref{sec:Pandk})
  \item If $k$ is large compared to $n$, we may not use permutation. Without permutation, the optimal condition for the PSS is derived. Also, the PSS using the rows of the cyclic Hadamard matrix provides the near-optimal PAPR reduction performance. (subSection \ref{sec:psswoP})
  \item With permutation, the optimal condition for the PSS is also derived, and the randomly generated PSS is the sub-optimal solution. (subSection \ref{sec:PSSwithP})
  \item The optimal condition for permutation functions is derived, where using the randomly generated permutation functions is the sub-optimal solution. (subSection \ref{sec:conforP})
  \item The derivation of the optimal conditions for PSS and permutation functions is meaningful because in practical systems using the deterministic PSS and permutation functions according to the derived optimal conditions is necessary because of memory saving and PAPR reduction performance guarantee.
\end{itemize}
In an attempt to further enhance the PAPR reduction performance in
the SLM scheme, future research directions will involve the systematic construction of PSS and permutation functions satisfying the derived optimal conditions by using \textit{sequence theory}. However, we defer it as a future work because it is out of scope of our goal.

\section{Simulation Results}
In this section, simulation results are presented to support the remarks in the previous section.
Here, the following simulation parameters are used in all plots; $N=64$, $M=4$ (quadrature phase shift keying (QPSK) is used), $n=16$, and $G=4$, which is reasonable because the OFDM-IM system has an advantage over the classical OFDM when they operate at low bit rates \cite{basar2017index}. CCDF is used to quantify the PAPR reduction performance. We performed $5\times 10^7$ trials by generating independent OFDM-IM blocks to plot each curve.
In the legend, ``perm.'' denotes the permutation.

\begin{figure}
  \centering
  \includegraphics[width=\linewidth]{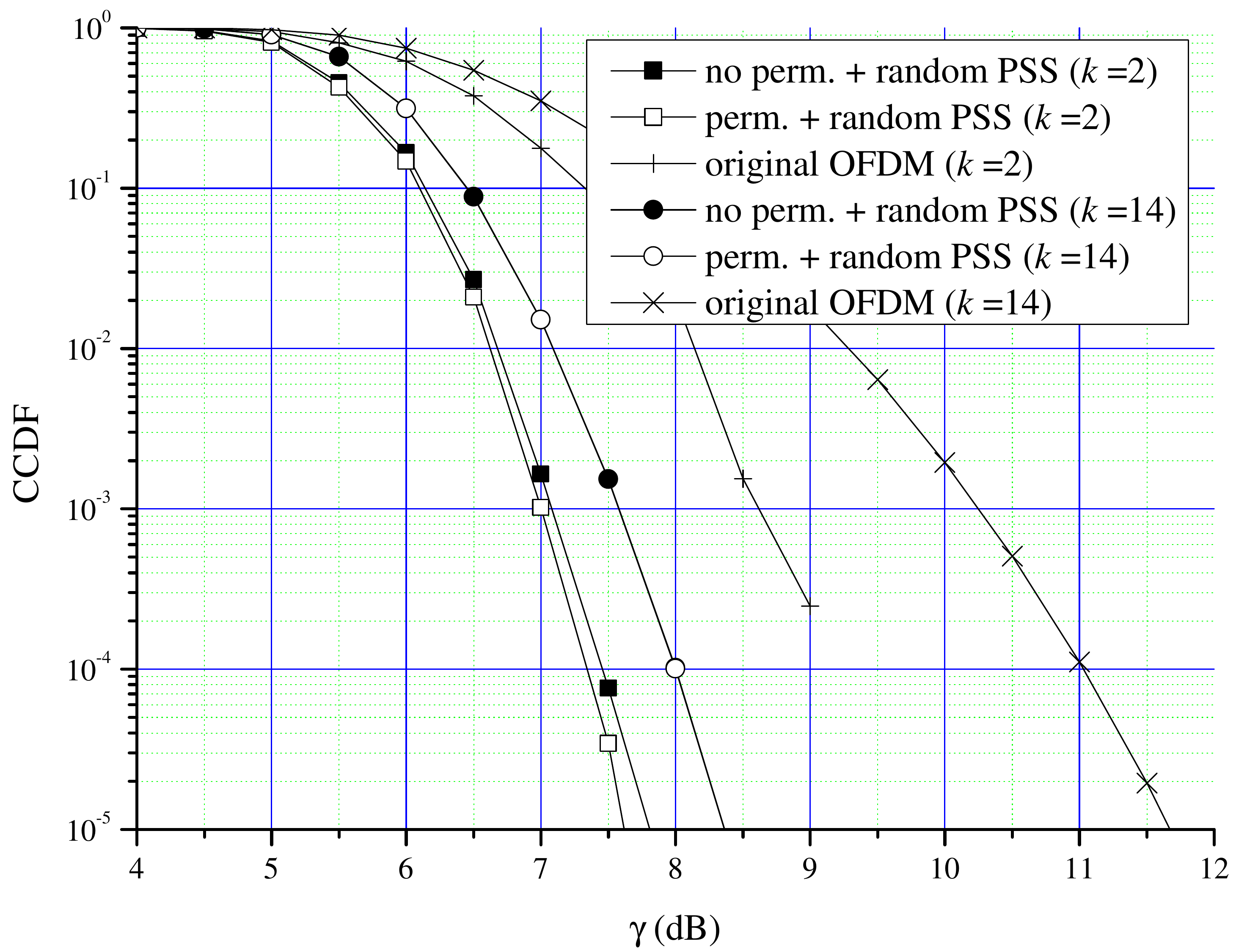}
  \caption{The PAPR reduction performances of the SLM scheme with $U=4$. The efficiency of permutation is valid if $k$ is small compared to $n$.}\label{fig:EffP}
\end{figure}
Fig. \ref{fig:EffP} depicts the CCDFs of OFDM-IM systems using the SLM scheme with $U=4$ and the randomly generated PSS is used. Two curves of original OFDM without SLM for $k=2$ and $k=14$ are shown as benchmark. Although the number of trials is large enough, the original OFDM case of $k=2$ has a rough appearance and there is no curve over PAPR = 9 dB. This is because the number of active subcarriers in an OFDM-IM block is only $K = 8$ in this case.
In Fig. \ref{fig:EffP}, for small $k$ compared to $n$, such as $k=2$, the permutation can improve the PAPR reduction performance over no permutation case while the gain of permutation is negligible for large $k$, analyzed in subSection \ref{sec:Pandk}. Due to lack of space, the plots of the other $k$'s are not depicted but they show the identical tendency.

\begin{figure}
  \centering
  \includegraphics[width=\linewidth]{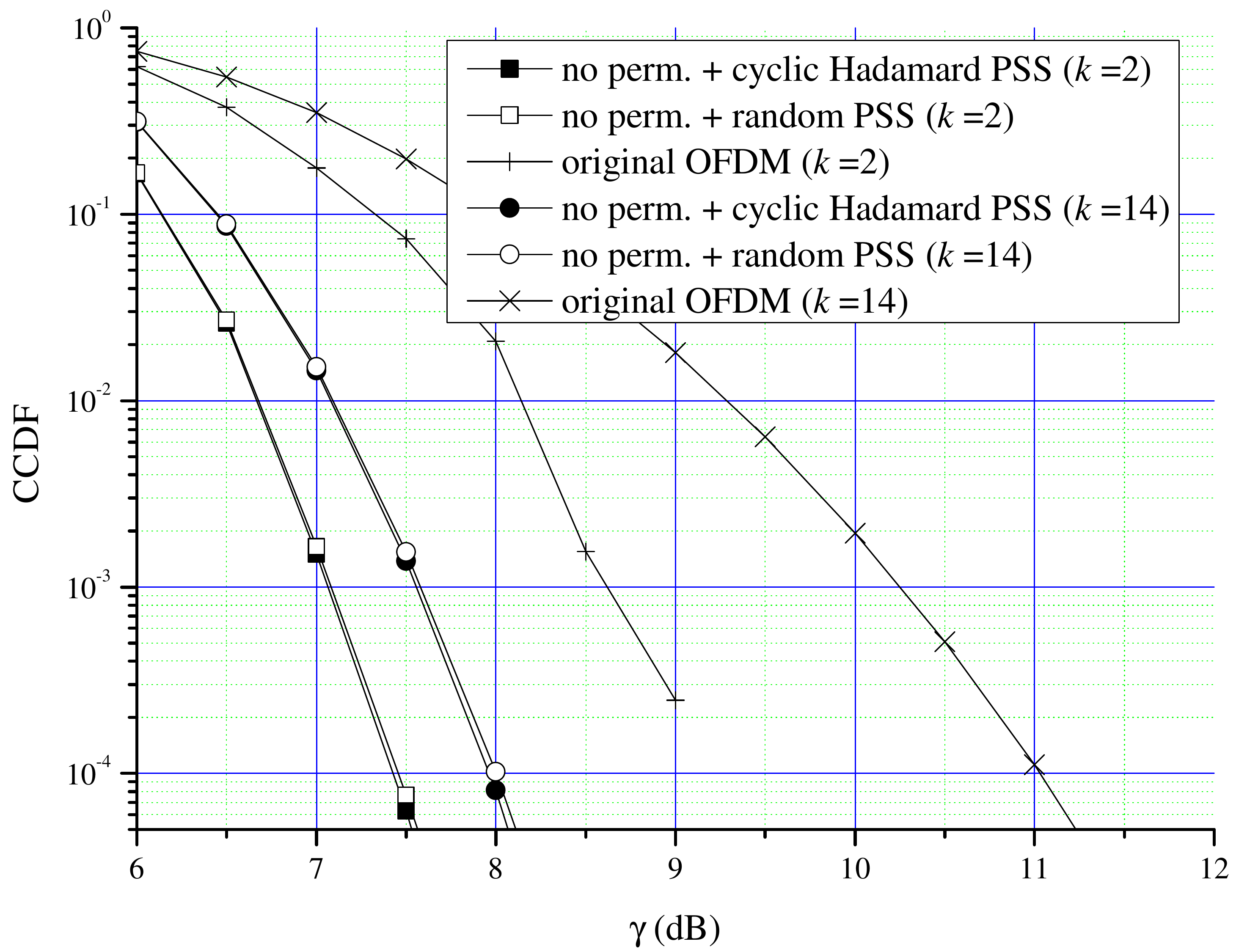}
  \caption{The PAPR reduction performances of the SLM scheme with $U=4$. Without permutation, the PSS constructed from the cyclic Hadamard matrix provides a better PAPR reduction performance over the random PSS case.}\label{fig:mseq}
\end{figure}
Fig. \ref{fig:mseq} depicts the CCDFs of OFDM-IM systems using the SLM scheme with $U=4$ and permutation procedure is not used.
As seen from Fig. \ref{fig:mseq}, without permutation, the PSS using the rows of the cyclic Hadamard matrix obviously provides the superior performance over the case using the random PSS as analyzed in subSection \ref{sec:psswoP}.

\begin{figure}
  \centering
  \includegraphics[width=\linewidth]{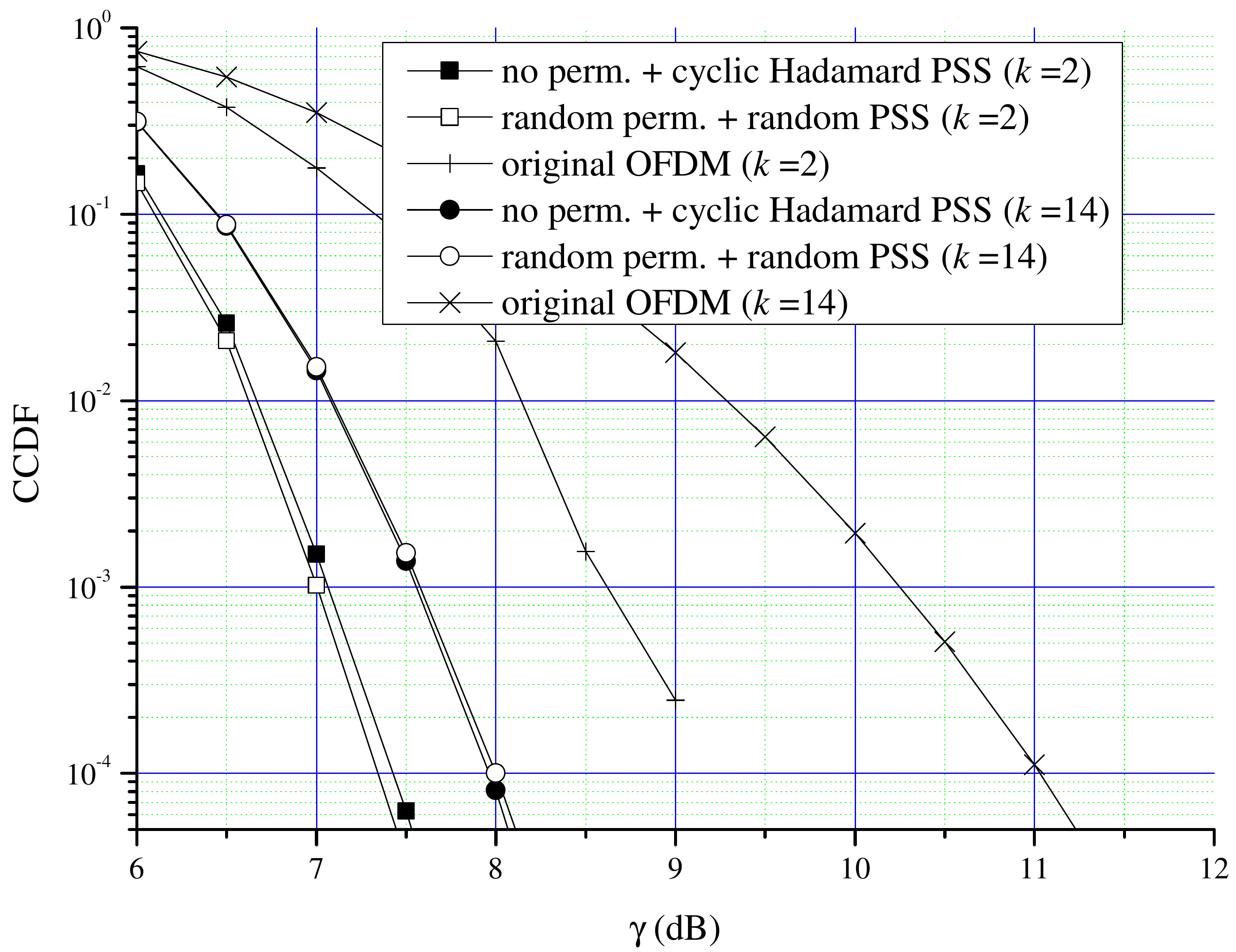}
  \caption{The PAPR reduction performances of the SLM scheme with $U=4$.}\label{fig:PvsnoP}
\end{figure}
Fig. \ref{fig:PvsnoP} depicts the CCDFs of OFDM-IM systems using the SLM scheme with $U=4$ and two original OFDM systems without SLM.
In Fig. \ref{fig:PvsnoP}, we can conclude that the permutation is not necessary if $k$ is large compared to $n$. There are two reasons as we already mentioned;
First, in the SLM scheme without permutation, the PSS can be near-optimally constructed from the cyclic Hadamard matrix. Second, the efficiency of permutation is negligible when $k$ is large.

\begin{figure}
  \centering
  \includegraphics[width=\linewidth]{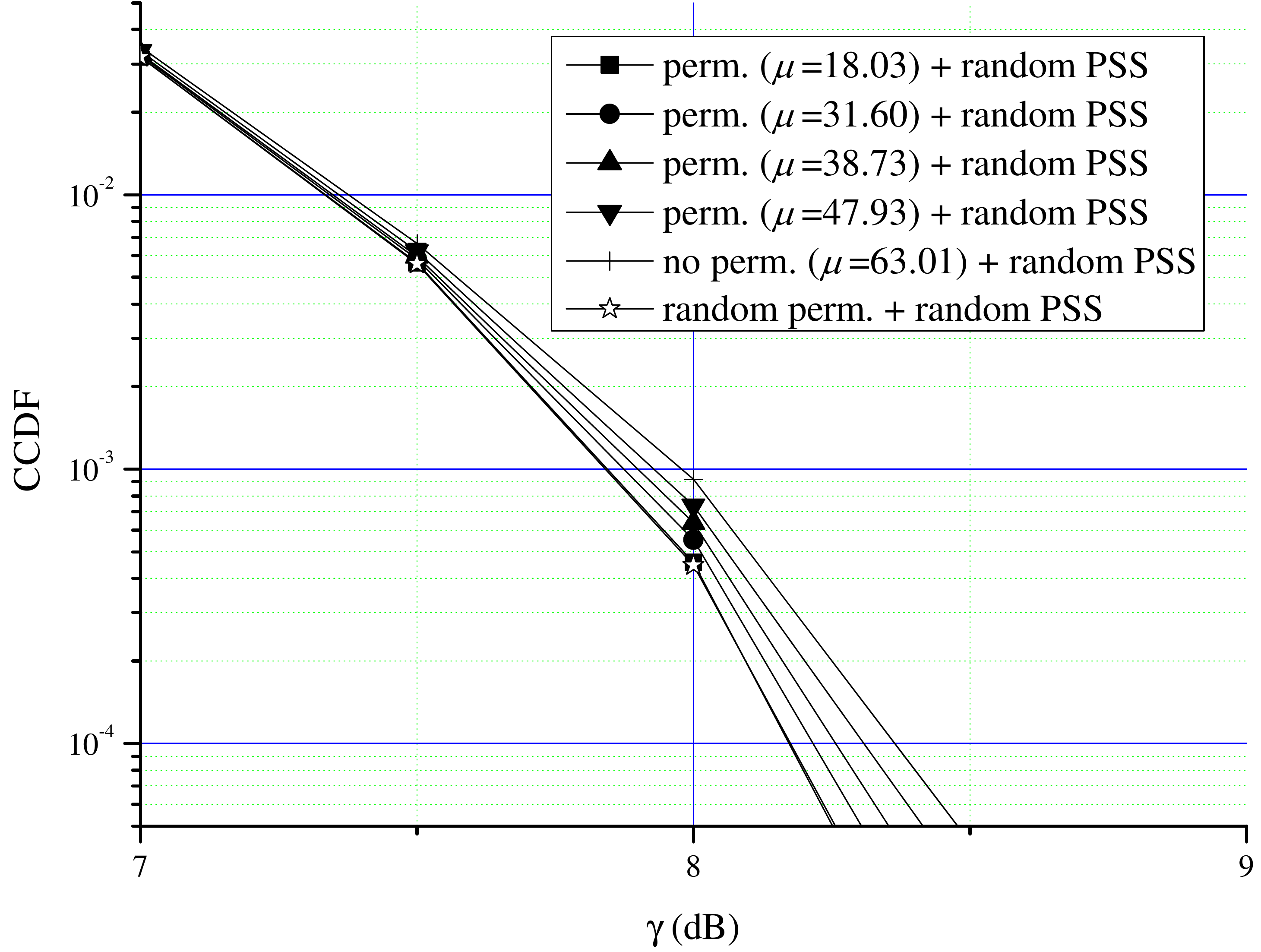}
  \caption{The PAPR reduction performances of the SLM scheme with $U=2$ in OFDM-IM systems with $k=2$. The permutation functions are manually generated according to various $\mu$'s.}\label{fig:var}
\end{figure}
Fig. \ref{fig:var} depicts the CCDFs of OFDM-IM systems with $k=2$ using the SLM scheme with $U=2$. The randomly generated PSS is used for all curves.
We denote the variance of the concatenated version of (\ref{eq:permdesign}) for all $l$'s as $\mu$ and generate the permutation functions having various $\mu$'s. Through massive simulations, the empirically best permutation functions are obtained with $\mu=18.03$. Note that no permutation case has the largest value of $\mu=63.01$.

Fig. \ref{fig:var} shows that the permutation functions with small $\mu$ are more appropriate for PAPR reduction. Thus, we confirm the validness of the optimal condition for the permutation functions derived in subSection \ref{sec:conforP}. As we expected, the randomly generated permutation functions provide almost the same PAPR reduction performance as the empirically best case of $\mu=18.03$. This is because the randomly generated permutation functions can lower the value of $\mu$ with high probability. However, using the deterministic permutation functions according to the derived optimal condition is more appropriate for practical situations from the view point of memory saving and performance guarantee.

\section{Conclusions}
OFDM-IM has become a promising technique whereby the specific activation of the frequency domain subcarriers is used for implicitly conveying extra information. Unfortunately, OFDM-IM has inherited a large PAPR problem from the classical OFDM system. To solve the large PAPR problem, SLM scheme is one of the promising schemes, but there are no works to analyze the optimal condition for phase sequences and permutation functions in the SLM scheme for OFDM-IM systems.

In this paper, useful aspects of the SLM scheme for OFDM-IM systems are described. First, there is a benefit of introducing the permutation procedure in the SLM scheme only if $k$ is small compared to $n$.
Therefore, if $k$ is large, the SLM scheme without permutation procedure could be a good choice with the PSS using the rows of the cyclic Hadamard matrix. Also, the optimal conditions for PSS and permutation functions in the SLM scheme are derived. These optimal conditions can guarantee the PAPR reduction performance of the SLM schemes, which is meaningful in practical systems. In order to support the analytical results, simulation results are presented.

\section*{Acknowledgment}
This work was supported by the Soonchunhyang University
Research Fund and the National Research Foundation of Korea (NRF) grant funded by the Korea government (MSIP; Ministry of Science, ICT $\&$ Future Planning) (No. NRF-2018R1C1B5030493).

\bibliographystyle{IEEEtran}
\bibliography{biblio,IEEEfull}

\end{document}